\documentclass[useAMS,usenatbib]{mn2e}
\usepackage{amsmath,amsfonts,epsfig,natbib,mathptm}
\def\be{\begin{equation}} 
\def\ee{\end{equation}} 
\def\ba{\begin{eqnarray}} 
\def\ea{\end{eqnarray}}

\def\reff@jnl#1{{\rm#1\/}}
\def\aj{\reff@jnl{AJ}}                  
\def\araa{\reff@jnl{ARA\&A}}            
\def\apj{\reff@jnl{ApJ}}                        
\def\apjl{\reff@jnl{ApJ}}               
\def\apjs{\reff@jnl{ApJS}}              
\def\ao{\reff@jnl{Appl.Optics}}         
\def\apss{\reff@jnl{Ap\&SS}}            
\def\aap{\reff@jnl{A\&A}}               
\def\aapr{\reff@jnl{A\&A~Rev.}}         
\def\aaps{\reff@jnl{A\&AS}}             
\def\azh{\reff@jnl{AZh}}                        
\def\baas{\reff@jnl{BAAS}}              
\def\jrasc{\reff@jnl{JRASC}}            
\def\memras{\reff@jnl{MmRAS}}           
\def\mnras{\reff@jnl{MNRAS}}            
\def\pra{\reff@jnl{Phys.Rev.A}}         
\def\prb{\reff@jnl{Phys.Rev.B}}         
\def\prc{\reff@jnl{Phys.Rev.C}}         
\def\prd{\reff@jnl{Phys.Rev.D}}         
\def\prl{\reff@jnl{Phys.Rev.Lett}}      
\def\pasp{\reff@jnl{PASP}}              
\def\pasj{\reff@jnl{PASJ}}              
\def\qjras{\reff@jnl{QJRAS}}            
\def\skytel{\reff@jnl{S\&T}}            
\def\solphys{\reff@jnl{Solar~Phys.}}    
\def\sovast{\reff@jnl{Soviet~Ast.}}     
\def\ssr{\reff@jnl{Space~Sci.Rev.}}     
\def\zap{\reff@jnl{ZAp}}                        
\def\nat{\reff@jnl{Nature}}             

\def\refn#1 {\noindent \hangindent=1em \hangafter=1 {#1} \par}

\title[A search for HCN at $z=4.695$]{Dense molecular gas in quasar host
galaxies: a search for HCN emission from BR B1202$-$0725 at $z = 4.695$}

\author[Isaak, Chandler \& Carilli]
{ Kate G. Isaak$^1$, Claire J. Chandler$^2$, Christopher L. Carilli$^2$\\
  $^1$ Astrophysics Group, Cavendish Laboratory, University of Cambridge, Cambridge CB3 0HE, UK\\
  $^2$ National Radio Astronomy Observatory, PO Box O, Socorro, NM 87801, USA\\
}
\pagerange{\pageref{firstpage}--\pageref{lastpage}}

\pubyear{2003}
\date{Accepted 2003 November 14.  Received 2003 November 7; in
original form 2003 September 15}

\begin{document}
\maketitle
\label{firstpage}

\begin{abstract}

We report on the results of a search using the VLA for redshifted
HCN(1--0) emission from the host galaxy of BR B1202$-$0725, an optically
luminous quasar at $z = 4.695$.  The host galaxy emits strongly in the
rest-frame far-infrared, and shows characteristics very similar to those
of more local ultraluminous infrared galaxies, in which a significant
fraction of the far-infrared emission is powered by star formation.
We find a 3-$\sigma$ upper limit to the HCN(1--0) emission of $4.9
\times 10^{10}$~K~km~s$^{-1}$~pc$^{2}$, assuming a $\Lambda$-cosmology.
This limit is consistent with correlations derived from measurements of
HCN, CO, and far-infrared emission for a sample of more local galaxies
including starbursts (Solomon et al.\ 1992a).

\end{abstract}


\section{Introduction}

Establishing the role of star formation in the early universe is key to
understanding the formation and evolution of galaxies.  Early searches for
the fingerprints of star formation at high-redshift received a tremendous
boost with the identification of the IRAS source F10214$+$4724 as a
hyperluminous infrared galaxy at $z=2.286$.  Millimetre and submillimetre
observations pointed to the presence of $\sim 10^8$~M$_\odot$ of dust
(Clements et al.\ 1992), and a molecular gas mass of $\sim 10^{10}$
(Solomon, Downes \& Radford 1992b).  Subsequent observations and analysis
showed that 10214+4724 was gravitationally lensed (e.g., Downes, Solomon
\& Radford 1995).  Nonetheless, F10214+4724 remains a ultraluminous
infrared galaxy (ULIRG), most likely undergoing one of its first massive
bursts of star formation.

ULIRGs represent a remarkable class of object, emitting a significant
fraction of their considerable bolometric luminosity at infrared
wavelengths.  With $L_{\rm FIR} > 10^{12}$~L$_{\odot}$ there is increasing
evidence to suggest that a large fraction of ULIRGs have recently been
in, or are currently undergoing, some sort of merger or galaxy-galaxy
interaction (e.g., Sanders \& Mirabel 1996).  These tidal interactions
are believed to trigger both extreme bursts of star formation and
the turn-on of a central active galactic nucleus (AGN), both of which
are observed, often simultaneously, in ULIRGs.  Local ULIRGs have been
studied extensively at millimetre/submillimetre wavelengths to assess the
relative importance of star formation, primarily through measurements of
the molecular gas (e.g., Solomon, Downes \& Radford 1992a; Solomon et
al.\ 1997; Gao 1996).  Important results have come from these studies:
in spite of the extreme CO luminosities observed in ULIRGs, the ratio
of $L_{\rm FIR}/L_{\rm CO}$ is an order of magnitude larger than normal
spiral galaxies; also, measurements of the dense gas tracer HCN(1--0)
show that a considerable fraction of the molecular gas in ULIRGs has
densities commonly observed in star forming cores.  These two results
can be interpreted as  high star formation rate per unit molecular gas
mass, and a high star formation potential.  The discovery of ULIRGs
at early epochs thus raises the question of whether star formation in
high-redshift sources is similar to that seen in local analogues?

The current generation of millimetre/submillimetre cameras (MAMBO and
SCUBA operating at the IRAM-30m and the James Clerk Maxwell Telescopes
respectively) has identified many candidate high-redshift ULIRGs, both
through targeted observations of objects known to be at high-redshift
(e.g., Archibald et al.\ 2001; Carilli et al.\ 2001) and through
blank-sky surveys (e.g., Bertoldi et al.\ 2000; Hughes et al.\ 1998).
Optical studies have shown that low-redshift quasars are located in
massive host galaxies (e.g., Boyce et al.\ 1998; Pagani, Falomo \& Treves
2003).  By extension, high-redshift quasars provide a convenient means
by which to pinpoint very distant, massive galaxies.  Many high-redshift
quasar host galaxies have been detected in the (sub)mm continuum (e.g.,
McMahon et al.\ 1994; Barvainis \& Ivison 2002; Omont et al.\ 2003;
Bertoldi et al.\ 2003), indicating the presence of massive quantities of
dust $10^{8-9}$~M$_\odot$ at early epochs.  The interpretation of the
(sub)mm continuum observations is not without complications, however,
since the underlying source of the UV energy heating the dust can be
due to either AGN and/or starburst activity.  CO emission has also been
detected from a number of these sources (e.g., Ohta et al.\ 1996; Omont
et al.\ 1996; Carilli et al.\ 2002a), unveiling massive molecular gas
reservoirs of $10^{10-11}$~M$_\odot$, and providing supporting evidence
for massive starbursts at less than 10\% of the age of the Universe.

One of the most spectacular examples of this class of object is BR
B1202$-$0725, the most distant ($z=4.695$) and optically luminous ($M_{B}=
-$28.5) radio-quiet quasar of the BR(I) quasars (Storrie-Lombardi
et al.\ 1996).  The first $z > 4$ quasar to be detected at (sub)mm
wavelengths (McMahon et al.\ 1994; Isaak et al.\ 1994), BR B1202$-$0725
was found to contain $\sim10^9$~M$_\odot$ of dust at a temperature of
$\sim 50$~K\@.  CO observations traced more than $10^{11}$~M$_\odot$
of molecular gas (assuming locally-determined conversion factors).
A multi-line LVG analysis by Ohta et al.\ (1998) using a range of CO
transitions (CO(7--6): Omont et al.\ 1996; CO(5--4): Ohta et al.\ 1996;
Omont et al.\ 1996; and CO(2--1): Ohta et al.\ 1998) suggested that
the molecular gas density is $>10^4$~cm$^{-3}$.  This conclusion was
also reached by Carilli et al.\ (2002b) using data having a much higher
signal-to-noise ratio.  The modelling requires a number of assumptions,
and it is clear that a direct measure of the dense gas is desirable in
order to establish whether the host galaxies of high redshift quasars
are indeed the analogues of nearby ULIRGs.

With few exceptions, studies of the molecular interstellar medium in very
distant galaxies have used the CO rotational ladder.  With a critical
density of $>10^4$~cm$^{-3}$, the lowest of the HCN rotational line
traces molecular gas at a much higher density than the corresponding CO
transition.  In this paper we present a search for HCN(1--0) emission from
BR B1202$-$0725 using the Very Large Array (VLA) of the National Radio
Astronomy Observatory\footnote{The National Radio Astronomy Observatory is
a facility of the National Science Foundation operated under cooperative
agreement by Associated Universities, Inc.}, with the aim of providing an
independent measure of the dense gas component in its host galaxy, and
to enable a direct comparison with the local ULIRGs observed by Solomon
et al.\ (1992a).  The mean rest frequency of the HCN(1--0) triplet
is $\nu_{\rm rest} = 88.632$~GHz, which is redshifted to 15.563~GHz
in BR B1202$-$0725.  All quantities given here have been derived
using a $\Lambda$-cosmology, with $H_0 = 65$~km~s$^{-1}$~Mpc$^{-1}$.
For comparison, the same quantities derived for a flat, Einstein-de
Sitter cosmology with $H_0 = 50$~km~s$^{-1}$~Mpc$^{-1}$ are included
in parentheses.  These cosmologies give a luminosity distance $D_{\rm L}
= 46.6(39.6)$~Gpc for BR B1202$-$0725.

\section{Observations and Data Reduction}

The VLA observations were made on 2000 August 26, September 2--4, and in
2003 January 25--29.  In 2000 the VLA was in the compact D configuration,
while in 2003 it was in the DnC configuration.  The instantaneous
bandwidth of the VLA is only 50~MHz, tuneable in finite steps of 20 or
30~MHz.  We therefore chose a set-up for the local oscillators that most
closely centred the redshifted HCN(1--0) line in the 50~MHz bandwidth,
and covered the 50~MHz with 8 channels of 6.25~MHz (120~km~s$^{-1}$) with
both right and left circular polarizations.  The narrow total bandwidth
of the current 15~GHz receivers at the VLA esulted in the sensitivity
achieved the frequency of the redshifted HCN line being degraded by
a factor of about 2 relative to the centre of the band.  Antenna-based
complex gains were monitored every 15 minutes through observations of the
quasar PMN J1149$-$0740.  The bandpass and absolute flux density scale
were determined through observations of 3C273 and 3C286 respectively.
The total uncertainty in the flux density calibration is estimated to
be less than 5 per cent.

The tropospheric phase stability was relatively poor for the observations
made during 2000 late summer, and the data with particularly bad
phase coherence (typically those on the longest baselines in the D
configuration) have been edited.  The weather during 2003 January was
excellent.  The data were reduced and imaged using the Astronomical
Image Processing System (AIPS).  After combining the data from both
sets of observations, the resulting naturally-weighted synthesized beam
is $4.6 \times 4.4$~arcsec$^2$ at position angle $-60^\circ$.  The rms
noise per channel in the final image is 60~$\mu$Jy~beam$^{-1}$.

\section{Results and Discussion}

\begin{figure}
\centering
\psfig{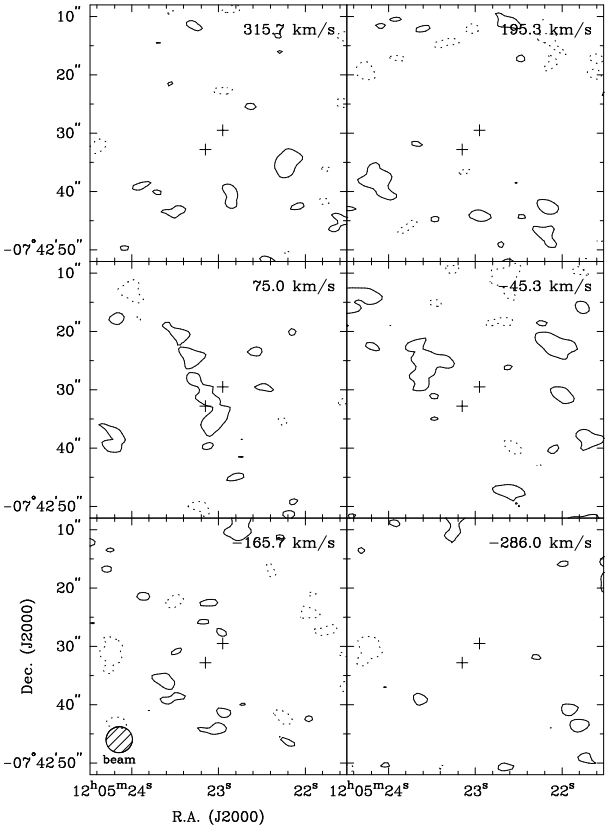}
\caption{Channel maps of the search for redshifted HCN(1--0) emission
from BR B1202$-$0725.  Contours are at 2-$\sigma$ intervals of
120~$\mu$Jy~beam$^{-1}$. The synthesized beam is given in the bottom
left panel.  The crosses denote the positions of the millimetre continuum
sources detected by Omont et al.\ (1996).  The velocity in the top right
of each panel is relative to a redshift $z = 4.695$.}
\label{map}
\end{figure}

No redshifted HCN(1--0) emission is detected.  The central six
channels are displayed in Fig.~\ref{map}, with 2-$\sigma$ contours
of 120~$\mu$Jy~beam$^{-1}$, and crosses denoting the positions of the
two millimetre sources detected by Omont et al.\ (1996).  The CO lines
reported by Omont et al.\ have line widths of 190 and 350~km~s$^{-1}$;
assuming an intrinsic line width $\Delta V_{\rm line}$ for any HCN
emission of $\sim 250$~km~s$^{-1}$ we find a 3-$\sigma$ upper limit to
the HCN(1--0) emission of ($3 \cdot (\Delta V_{\rm line}/\Delta V_{\rm
channel})^{1/2} \cdot$~rms per channel) = 31~mJy~km~s$^{-1}$.  Using
equations~1 and 3 from Solomon et al.\ (1992b), we derive a 3-$\sigma$
upper limit to the line luminosity of $L'_{\rm HCN} < 4.9(3.6) \times
10^{10}$~K~km~s$^{-1}$~pc$^{2}$.

\begin{figure}
\centering
\psfig{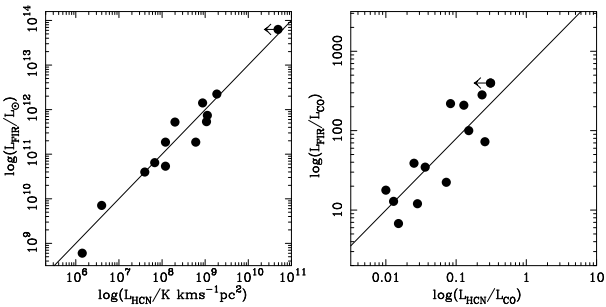}
\caption{{\it Left:} $L_{\rm FIR}$ vs.\ $L_{\rm HCN}$ for a sample
of galaxies, ranging from normal galaxies to those undergoing massive
starburst, from Solomon et al.\ (1992a).  The solid line represents a
simple least-squares fit to the Solomon et al.\ data points.  The arrow
at top right represents the 3-$\sigma$ upper limit to $L_{\rm HCN}$ for
BR B1202$-$0725 reported here.  {\it Right:} $L_{\rm FIR}$ vs.\ $L_{\rm
HCN}$ normalised by the CO luminosity, after Solomon et al.\ (1992a).}
\label{plots}
\end{figure}

In order to compare the molecular properties of BR B1202$-$0725 with
those of local (U)LIRGs we include it in plots of $L'_{\rm HCN}$ vs.\
$L_{\rm FIR}$, and $L_{\rm FIR}/L'_{\rm CO}$ vs.\ $L'_{\rm HCN}/L'_{\rm
CO}$, in Fig.~\ref{plots}.  The value for the far-infrared luminosity,
$L_{\rm FIR} = 6.3(4.6) \times 10^{13}$~L$_\odot$, has been calculated
by integrating under the thermal greybody spectral energy distribution
as detailed by Isaak et al.\ (2002).  The CO(1--0) luminosity comes from
a recent measurement of the CO(1--0) line flux, $S_{\rm CO} = 0.17 \pm
0.05$~Jy~km~s$^{-1}$ (C. Henkel, private communication).  The upper limit
of $L'_{\rm HCN}$ for BR B1202$-$0725 is consistent with the correlations
between $L'_{\rm HCN}$ vs.\ $L_{\rm FIR}$, and $L_{\rm FIR}/L'_{\rm CO}$
vs.\ $L'_{\rm HCN}/L'_{\rm CO}$, observed in more local (U)LIRGs.

Our measured upper limit suggests that the globally-averaged molecular
properties of the host galaxy of BR B1202$-$0725 are consistent with
those found in more local ULIRGs. Furthermore, it is consistent with
the interpretation that a large fraction of the far-infrared luminosity
originates from dust heated by star formation.  We note, however, that
a significant fraction of ULIRGs show signs of both starburst and AGN
activity (Genzel et al.\ 1998).  This has also been found to be the case
in many of the high-redshift sources identified in submillimetre surveys
that have exploited the lensing effect of galaxy clusters (e.g., Smail,
Ivison \& Blain 1997).  It is therefore not surprising that a naked AGN
residing in a young host galaxy might show similarities with ULIRGs.

A deeper limit to the HCN emission from BR B1202$-$0725 will only be
possible with the EVLA\footnote{Details of the VLA expansion project
may be found at http://www.aoc.nrao.edu/evla/}, when an improvement in
sensitivity of a factor of about 6 is anticipated for spectral line work
at this frequency. With our current limit we do, however, exclude a low
$L_{\rm FIR}/L_{\rm HCN}$ ratio.  Such a ratio would be interpreted as
a host galaxy with a high molecular gas density with little ongoing star
formation or AGN activity, that is, gas that is {\it about} to form stars
or be accreted onto a central black hole.  A high $L_{\rm FIR}/L_{\rm
HCN}$ ratio is still consistent with the measured upper limit, and
would indicate that a significant contribution is made to $L_{\rm FIR}$
from dust heated by the central AGN, rather than by star formation.
A global measure of the HCN emission from a galaxy in which both AGN and
starburst activity are observed cannot definitively establish the dominant
source of the far-infrared luminosity, as strong HCN emission is seen in
both starbursts and local AGN (Seyfert galaxies: Curran et al.\ 2001).
Differentiating between the two may be possible with images of both
considerably higher angular resolution and of higher HCN transitions.
In this way it should be possible to trace the distribution of the denser
gas as well as its excitation.

\section*{ACKNOWLEDGEMENTS}

KGI acknowledges support of a PPARC fellowship and the Cavendish
Astrophysics Group. We thank the anonymous referee for comments that
helped to clarify points made in this paper.  This research has made
use of NASA's Astrophysics Data System.


\section*{REFERENCES}

\refn{Archibald E. N., Dunlop J. S., Hughes D. H., Rawlings S., Eales
S. A., Ivison R. J., 2001, MNRAS, 323, 417}
\refn{Barvainis R., Ivison R., 2002, ApJ, 571, 712}
\refn{Bertoldi F., Carilli C. L., Cox P., Fan X., Strauss M. A.,
Beelen A., Omont A., Zylka R., 2003, A\&A, 406, L55}
\refn{Bertoldi F. et al., 2000, A\&A, 360, 92}
\refn{Boyce P. J. et al., 1998, MNRAS, 298, 121}
\refn{Carilli C. L. et al., 2001, ApJ, 555, 625}
\refn{Carilli C. L. et al., 2002a, ApJ, 575, 145}
\refn{Carilli C. L. et al., 2002b, AJ, 123, 1838}
\refn{Clements D. L., Rowan-Robinson M., Lawrence A., Broadhurst T.,
McMahon R., 1992, MNRAS, 256, 35P}
\refn{Curran S. J., Polatidis A. G., Aalto, S., Booth R. S., 2001,
A\&A, 373, 459}
\refn{Downes D., Solomon P. M., Radford S. J. E., 1995, ApJ, 453, L65}
\refn{Gao Y., 1996, PhD Thesis, SUNY at Stony Brook}
\refn{Genzel R. et al., 1998, ApJ, 498, 579}
\refn{Hughes D. H. et al., 1998, Nature, 394, 241}
\refn{Isaak K. G., McMahon R. G., Hills R. E., Withington S., 1994,
MNRAS, 269, L28}
\refn{Isaak K. G., Priddey R. S., McMahon R. G., Omont A., Peroux C.,
Sharp R. G., Withington S., 2002, MNRAS, 329, 149}
\refn{McMahon R. G., Omont A., Bergeron J., Kreysa E., Haslam C. G.
T., 1994, MNRAS, 267, L9}
\refn{Ohta K., Nakanishi K., Akiyama M., Yamada T., Kohno K., Kawabe
R., Kuno N., Nakai N., 1998, PASJ, 50, 303}
\refn{Ohta K., Yamada T., Nakanishi K., Kohno K., Akiyama M., Kawabe
R., 1996, Nature, 382, 426}
\refn{Omont A., Beelen A., Bertoldi F., Cox P., Carilli C. L.,
Priddey R. S., McMahon R. G., Isaak K. G., 2003, A\&A, 398, 857}
\refn{Omont A., Petitjean P., Guilloteau S., McMahon R. G., Solomon P.
M., Pecontal E., 1996, Nature, 382, 428}
\refn{Pagani C., Falomo R., Treves A., 2003, ApJ, 596, 830}
\refn{Sanders, D. B., Mirabel, I. F., 1996, ARA\&A, 34, 749}
\refn{Smail I., Ivison R. J., Blain A. W., 1997, ApJ, 490, L5}
\refn{Solomon P. M., Downes D., Radford S. J. E., 1992a, ApJ, 387, L55}
\refn{Solomon P. M., Downes D., Radford S. J. E., 1992b, ApJ, 398, L29}
\refn{Solomon P. M., Downes D., Radford S. J. E., Barrett J. W.,
1997, ApJ, 478, 144}
\refn{Storrie-Lombardi L. J., McMahon R. G., Irwin M. J., Hazard C.,
1996, ApJ, 468, 121}

\label{lastpage}
\bsp

\end{document}